\documentclass[aps,prl,groupedaddress,showpacs,preprint]{revtex4}

\usepackage{epsfig}
\usepackage{graphicx}
\usepackage{dcolumn}
\usepackage{amsmath}

\begin{document}
\bibliographystyle{apsrev}

\title{Stabilization of amorphous GaN by oxygen}

\author{F. Budde}

\author{B.J. Ruck}

\author{A. Koo}

\author{S. Granville}

\author{H.J. Trodahl}

\affiliation{MacDiarmid Institute for Advanced Materials and
Nanotechnology, School of Chemical and Physical Sciences, Victoria
University of Wellington, P.O. Box 600, Wellington, New Zealand}

\author{A. Bittar}

\author{G.V.M. Williams}

\affiliation{Measurement Standards Laboratory, Industrial Research
Ltd., P.O. Box 31310, Wellington, New Zealand}

\author{M.J. Ariza}

\affiliation{Grupo de F\'{i}sica de Fluidos Complejos,
Departamento de F\'{i}sica Aplicada Universidad de Almer\'{i}a,
E-04120 Almer\'{i}a, Spain}

\author{B. Bonnet}

\author{D.J. Jones}

\affiliation{Laboratoire des Agr\'{e}gats Mol\'{e}culaires et
Mat\'{e}riaux Inorganiques UMR CNRS 5072, Universit\'{e}
Montpellier II, Place Eug\`{e}ne Bataillon F-34095 Montpellier
C\'{e}dex 5, France}

\author{J.B. Metson}

\affiliation{Department of Chemistry, Auckland University, Private
Bag 92019, Auckland, New Zealand}

\author{S. Rubanov}

\author{P. Munroe}

\affiliation{Electron Microscope Unit, University of New South
Wales, Sydney, NSW 2052, Australia}

\date{\today}

\begin{abstract}
Ion assisted deposition (IAD) has been investigated for the growth
of GaN, and the resulting films studied by x-ray diffraction and
absorption spectroscopy and by transmission electron microscopy.
IAD grown stoichiometric GaN consists of random-stacked
quasicrystals of some 3 nm diameter. Amorphous material is formed
only by incorporation of 15\% or more oxygen, which we
attribute to the presence of non-tetrahedral bonds centered on
oxygen. The ionic favorability of heteropolar bonds and its
strikingly simple constraint to even-membered rings is the likely
cause of the instability of stoichiometric a-GaN.
\end{abstract}


\maketitle
\flushbottom

The existence and stability  of glasses and amorphous solids is
poorly understood in detail~\cite{Angell,Sokolov}, despite
extensive technological applications that exploit them. The
problem is exacerbated by their metastability; they can be formed
only in non-equilibrium processes and their bonding configuration
depends on the details of that process. At the most stable extreme
there are covalently bonded materials which form continuous random
networks in a simple quench from a melt, of which the two-fold
bonded chalcogenide atoms are often a constituent.
Tetrahedrally-bonded materials require more rapid quenching, so
that amorphous Si (a-Si) and Ge are formed by deposition from the
vapor onto an ambient-temperature substrate, or by ion
implantation or bombardment~\cite{Mott_Davis}. In the case of the
III-V compounds, including
GaN~\cite{Bittar_Markwitz__Koo_Trodahl2,Hariu_Shibata,Kuball,Hassan,Chen_Kordesch,Kang,Nonomura,Miyazaki_Ohtsuka},
claims have been made for amorphous-material preparation by the
same techniques as are used in Si technology, though these
materials have not been studied at the same level as a-Si. We will
argue below that in the case of GaN, at least, the truly amorphous
form is unstable in stoichiometric material.

There is ambiguity in characterizing a material as amorphous, for
the difference between nanocrystalline and amorphous materials is
one of the range over which atomic order persists. It is
convenient to consider Si as a model for tetrahedrally-bonded
amorphous materials. In its amorphous form it consists of nearly
ideally tetrahedrally bonded Si atoms with almost completely
random dihedral angles~\cite{Mott_Davis,Zallen}. That structure
leads to first and second nearest neighbor shells with the same
coordination and at very similar distances as are found in
crystalline Si, but there remains almost no clear order at larger
distances. The amorphous form of a covalently bonded material is
then characterized by a rapid loss of configurational order at
distances beyond that at which the covalent bonding firmly
constrains the configuration.

A-Si solar cell and thin-film transistor technologies have now led
to a substantial industry, but that success has not been  matched
in the amorphous group III-V compounds. However, it has been
suggested that the amorphous form of GaN (a-GaN) might be
particularly interesting in this regard, because the relatively
strong ionicity of the Ga-N bond would be expected to limit
homopolar bonds that might otherwise introduce states in the gap
separating the valence and conduction bands~\cite{Stumm_Drabold}.
There have been several attempts to form and study thin films of
the material, with mixed results as regards both its preparation
and electronic and optoelectronic
properties~\cite{Bittar_Markwitz__Koo_Trodahl2,Hariu_Shibata,Kuball,Hassan,Chen_Kordesch,Kang,Nonomura,Miyazaki_Ohtsuka}. Many
of the attempts have generated either nanocrystalline or
nonstoichiometric material, and it is unclear from a study of the
literature that genuinely stoichiometric a-GaN can be
formed.

The case of tetrahedrally bonded III-V compounds is also of
special fundamental interest, for these materials add one further
constraint to those in Si, namely the need to accommodate
predominantly heteropolar bonds. It is important in this regard to
note that the most successful structural models for a-Si contain
odd-membered rings of atoms~\cite{Polk}, and those would require
at least one homopolar bond and are thus formed only at a
substantial energy cost in the III-Vs. Thus extensive extended
x-ray absorption fine structure (EXAFS) studies of amorphous
III-As and -P films formed by ion bombardment have concluded that
these necessarily contain homopolar
bonds~\cite{Glover_Yoon__Ridgway_Foran}. The nitrides are
especially interesting in this regard, for they are the most ionic
of the III-V compounds and can be expected to be the least
tolerant of homopolar bonds. Molecular dynamic studies suggest
that these materials might still form an amorphous structure, but
that they contain four- as well as six-membered rings, a wide
range (85$^\circ$--130$^\circ$) of Ga-N-Ga and N-Ga-N bond angles,
and a large fraction of triply coordinated ions with sp$^2$
bonds~\cite{Chen_Drabold}. Such a structure is expected to display
specific signatures, most clearly as regards the presence of
sp$^2$ bonded ions. Thus, in addition to their potential
optoelectronic use, these materials offer the possibility of
determining the effects of an extra constraint on the stability of
the amorphous phase.

We have developed an ion-assisted deposition technique for the
preparation of
GaN~\cite{Bittar_Markwitz__Koo_Trodahl2,Koo_Trodahl1,Lanke_Trodahl}. The
materials are grown in a vacuum system with a base pressure of
$5\cdot10^{-9}$~mbar using a Ga evaporation source and a
Kaufmann-type ion source. The composition of the resulting films
has been characterized by Rutherford backscattering spectroscopy
(RBS), and we find that the films are Ga rich for low ion energy,
rising to stoichiometric GaN when the ion energy is above 300
eV~\cite{Lanke_Trodahl}. At these energies the N$_2^+$ penetration
depth is about 0.5 nm, which disturbs the top two atomic layers
and encourages a reconstruction of the bonds; the preparation thus
involves a deposit of energy intermediate between that of
evaporated vapor deposition and ion bombardment. The oxygen
concentration in the films is less than 1 atomic \%, but can be
increased to over 20\% by adding a water vapor partial
pressure of $4\cdot10^{-6}$~mbar. The oxygen appears to substitute
for nitrogen, and excess N is trapped within the network as
molecules~\cite{Metson_Bittar,Ruck_Markwitz}. We now report x-ray
diffraction (XRD), Ga-edge EXAFS, and transmission electron
microscopy (TEM), all of which yield information about the atomic
configurational order in the films.

\begin{figure}
  \centering
  \includegraphics[width=6.5cm,keepaspectratio]{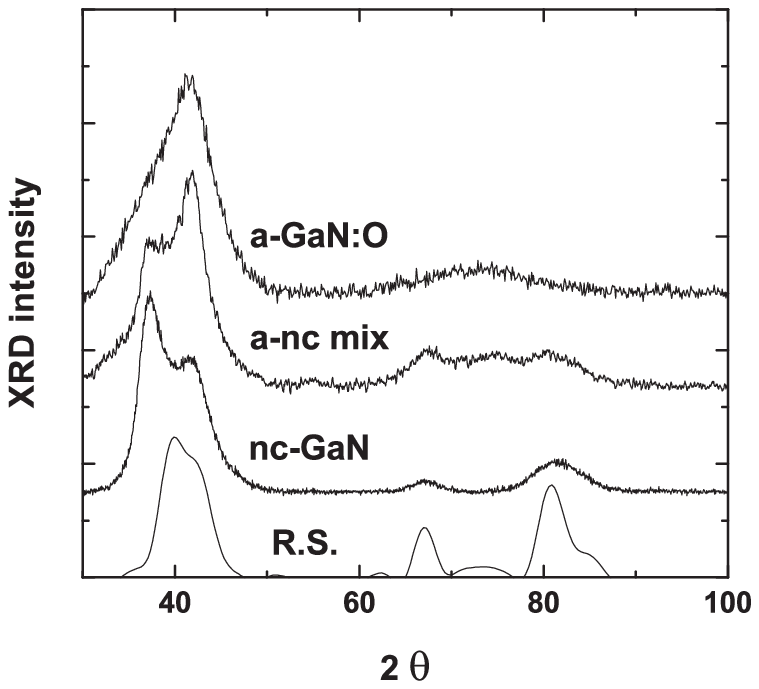}
  \caption{XRD patterns from stoichiometric nanocrystalline (nc) GaN, an
amorphous (a) film with 23\% O and a mixed nc-a film. R.S.~is
a simulated pattern for random-stacked quasicrystals with an
average diameter of 3 nm. The text discusses the 36$^{\circ}$ peak
for nc material which is not reproduced by the simulation.}
  \label{Fig1}
\end{figure}

In earlier publications we reported that the procedure leads to
amorphous GaN containing 10--20\% O in films prepared in a
base pressure of $5\cdot10^{-6}$ mbar (mainly water
vapor)~\cite{Bittar_Markwitz__Koo_Trodahl2}. In contrast, our
near-stoichiometric films are nanocrystalline, as demonstrated by
the XRD data in Figure~\ref{Fig1}. The bottom trace, taken on a
film with little oxygen, shows a clear diffraction pattern, with a
broadening corresponding to configurational order extending to
only about 3 nm. Crystalline material with similar dimensions are
also seen to fill high resolution TEM images of these films.
Interestingly, the XRD features do not agree with either the
wurtzite nor the zincblende phases of GaN, but they show
substantial agreement with the synthetic pattern labelled R.S. in
the figure, a simulation for 3 nm diameter random-stacked
quasicrystals. The Ga-Ga separation for the simulation was 0.324
nm, about 2\% larger than the separations in either wurtzite or
zincblende GaN. Scattering from N atoms and both dynamic and
static short-range positional disorder (the Debye-Waller factor)
are ignored in the simulation, which cannot then be expected to
reproduce the magnitudes of the diffraction peaks faithfully, but
with one exception the peak positions fit very well. That feature,
at 36$^{\circ}$ in the pattern, may indicate that some of the
nanocrystals have the zincblende structure~\cite{Kim_Holloway}. It
is significant in this context that oriented films with very high
densities of stacking faults, approaching in some cases the random
stacked density, have been formed under a variety of deposition
conditions~\cite{Siegle_Lischka,Yang_Ploog,Munkholm_Speck}. The
upper spectrum in Fig.~\ref{Fig1} shows the XRD pattern for a film
containing 23\% oxygen, and the difference from
nanocrystallinity is striking. All evidence of crystalline order
is missing in material with oxygen concentration above about 15\%, at which every Ga atom might have one O in its first shell.
The middle pattern in Fig.~\ref{Fig1}, for a film with an
intermediate oxygen concentration, is a clear mixture of amorphous
and nanocrystalline. Again in these films the TEM images are in
agreement with an amorphous structure.

\begin{figure}
  \centering
  \includegraphics[width=6.5cm,keepaspectratio]{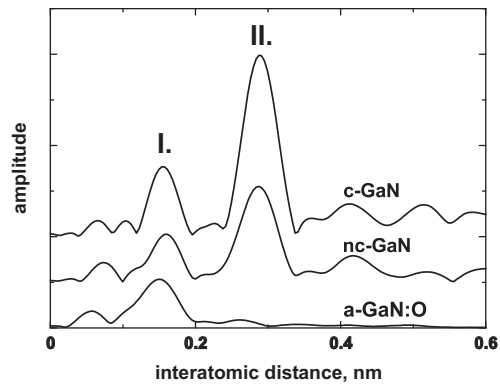}
  \caption{1st (\textsf{I.}) and 2nd (\textsf{II.}) neighbor peaks in the Fourier transformed EXAFS data for a
nanocrystalline and an amorphous film, compared with a crystalline
film.}
  \label{Fig2}
\end{figure}

These conclusions are corroborated by the Ga-edge EXAFS data in
Figure~\ref{Fig2}. The first neighbor peak in those results is
consistent with four nitrogen atoms at a similar distance to that
in crystalline GaN~\cite{Koo_Trodahl1}. It should be noted however
that our EXAFS data, which are not phase corrected, cannot
distinguish between backscattering from nitrogen or from oxygen
atoms. The second nearest neighbor peak shows marked differences
between nanocrystalline and amorphous films. Those films prepared
without oxygen are characterized by an intense second maximum
corresponding to backscattering from 12 gallium atoms. In
contrast, films containing oxygen show almost no second neighbor
peak, not unlike the EXAFS spectra of a-Si, a-Ge, and other
amorphous III-V
compounds~\cite{Glover_Yoon__Ridgway_Foran,Glover_Ridgway__Ridgway_NylandstedLarson2}.
The severe loss of intensity of this peak signals a very high
degree of static disorder.

\begin{figure}
  \centering
  \includegraphics[width=7.0cm,keepaspectratio]{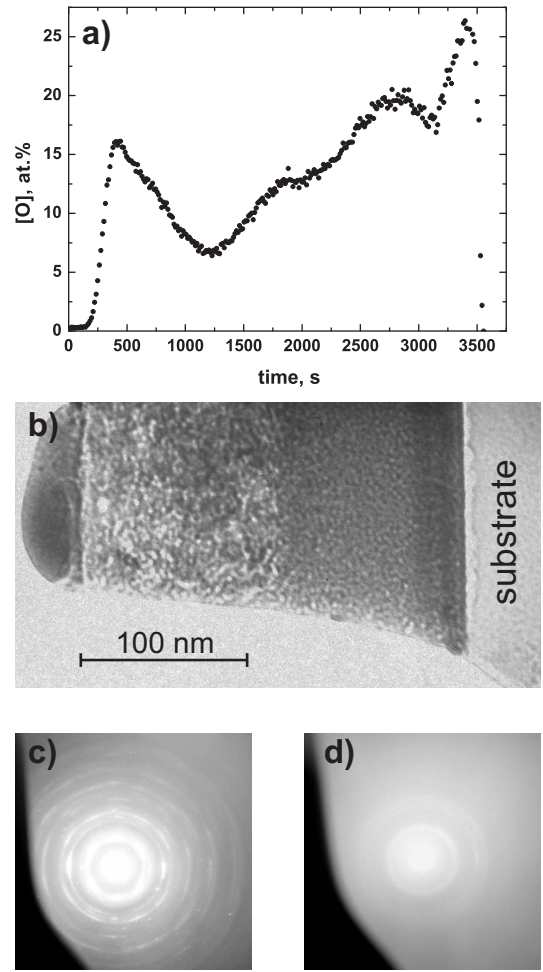}
  \caption{(a)~Oxygen concentration through a film
determined by SIMS (calibrated by RBS). (b)~TEM image of a film
with graded oxygen concentration. An interface running close to
the center, separating an amorphous oxygen-rich layer from a
nanocrystalline, is clearly visible. (c)~and (d) Selected area
electron diffraction patterns from the two layers.}
  \label{Fig3}
\end{figure}

The amorphous versus nanocrystalline transition is manifest in a
remarkably sudden function of the O concentration, as evidenced by
the TEM image and diffraction patterns in Figure~\ref{Fig3}. The
image is of a film prepared in a chamber in which a partial
pressure of water vapor continually dropped, so that the oxygen
concentration, determined by secondary ion mass spectroscopy
(SIMS) [Fig.~\ref{Fig3}(a)], fell from about 25\% at the
substrate to a minimum of 7\% before rising to 15\% toward
the surface in contact with air. Despite the gradual variation in
O concentration through the film, the image shows a sharp
separation where the structure changes from amorphous adjacent to
the substrate to nanocrystalline nearer to the outer surface.

The full body of evidence then raises an interesting question: Why
does GaN require impurities to form the amorphous phase? Why is
stoichiometric GaN so reluctant to form the amorphous phase, while
the similarly-bonded a-Si is formed so readily? The local bonding
order is identical, so one must look for an explanation beyond
that scale. The one strong symmetry lost in a- vs.\ c-Si is the
disordering of the dihedral angle, but the similar energies of
wurtzite and zincblende GaN and the prevalent random stacking
suggests that the dihedral angle is even more weakly constrained
than in Si. Evidently the heteropolar bonding and its implied
absence of odd-membered rings destabilizes a-GaN and encourages
the growth of nanocrystalline material. The conclusion agrees with
the presence of homo-polar bonds in a-III-V compounds involving
less strongly ionic bonds~\cite{Glover_Yoon__Ridgway_Foran}, and
with the predominance of three-fold bonds in molecular dynamics
simulations of a-GaN~\cite{Chen_Drabold}.

The description above suggests an obvious explanation for the
stabilization of the amorphous phase by O. The bonding
configuration of such high concentrations of a chalcogenide in
disordered GaN is unlikely to be exclusively tetrahedral, as
evidenced by observations that in high concentrations oxygen impurities in c-GaN reside
in dislocations and grain boundaries rather than as a simple
nitrogen substituent~\cite{Butcher_Freitas,Elsner_Briddon}. We
suggest that it forms either two- or four-fold bonds, primarily
with Ga ions, which reduces the density of exclusively tetrahedral
bonds and permits a space-filling network even within the
even-membered ring constraint. Alternatively there may be two- or
three-fold O ions forming bridges between Ga and N ions, reducing
the constraints both as regards odd-membered rings and
non-tetrahedral bonds.

These data raise the question as to which, if any, of these
configurations are suitable for optoelectronic devices. The strong
ionicity of the bonds in GaN that was at the core of the
suggestion that a-GaN might be a useful optoelectronic material is
itself the constraint that prevents the formation of a-GaN, at
least by the IAD techniques described in this paper. Optical
absorption data show quite clearly that the nanocrystalline films
have an absorption edge very close to the 3.4 eV that is found in
crystalline GaN, but there is a tail to lower energies indicating
the existence of states in the gap~\cite{Koo_Trodahl1}. The edge
is still reasonably sharp, so that the 200 nm thick films are
reasonable transparent across the visible. Amorphous GaN:O films
have a larger band gap and an edge which is as sharp as found in
nanocrystalline films, and as we have reported earlier they also
show a photoconductivity which is larger by several orders of
magnitude~\cite{Koo_Trodahl1}.

In summary we have identified conditions leading to stoichiometric GaN, but
those conditions lead only to a nanocrystalline structure, with
nearly random-stacked quasi-crystals of about 3 nm diameter. The
picture that emerges suggests that amorphous GaN is not as stable
as a-Si, despite the similarity of bonding in the two materials.
It is likely that the heteropolar character of the bonds in GaN,
which adds a further constraint preventing the formation of
odd-membered rings, is responsible for the apparent instability of
a-GaN. On the other hand, clearly amorphous films do form with the
introduction of oxygen into the bonding network, which we
associate with a relaxation of constraints on the bonding network
by the presence of non-tetrahedral bond configurations centered on
oxygen.

\begin{acknowledgments}
We acknowledge the New Zealand Foundation for Research, Science
and Technology for financial assistance through its New Economy
Research Fund, and through doctoral (AK) and postdoctoral (BJR)
fellowships and the Royal Society of New Zealand for support under
the Centre of Research Excellence programme.
\end{acknowledgments}


\clearpage

\end{document}